\documentclass[a4paper,fleqn,usenatbib,useAMS,letters]{mnras}
\pdfoutput=1
\usepackage{mathptmx}

\usepackage[T1]{fontenc}
\usepackage{ae,aecompl}

\usepackage{graphicx}	
\usepackage{amsmath}	
\usepackage{amssymb}	

\newcommand{\gtot}{\ensuremath{\mathrm{g}_{\mathrm{t}}}}
\newcommand{\gbar}{\ensuremath{\mathrm{g}_{\mathrm{b}}}}
\newcommand{\gD}{\ensuremath{\mathrm{g}_{\mathrm{D}}}}
\newcommand{\Mbar}{\ensuremath{\mathrm{M}_{\mathrm{b}}}}
\newcommand{\MD}{\ensuremath{\mathrm{M}_{\mathrm{D}}}}

\newcommand{\azero}{\ensuremath{\mathrm{a}_{0}}}

\title[Testing Verlinde's Emergent Gravity]{Testing Verlinde's Emergent Gravity with the Radial Acceleration Relation}

\author[F. Lelli et al.]{
Federico Lelli,$^{1}$\thanks{ESO Fellow. E-mail: flelli@eso.org}
Stacy S. McGaugh,$^{2}$
James M. Schombert$^{3}$
\\
$^{1}$European Southern Observatory, Karl-Schwarschild-Strasse 2, Garching bei M\"{u}nchen, Germany\\
$^{2}$Department of Astronomy, Case Western Reserve University, 10900 Euclid Avenue, Cleveland, OH 44106, USA\\
$^{3}$Department of Physics, University of Oregon, Eugene, OR 97403, USA
}

\date{Accepted XXX. Received YYY; in original form ZZZ}

\pubyear{2017}

\begin{document}
\label{firstpage}
\pagerange{\pageref{firstpage}--\pageref{lastpage}}
\maketitle

\begin{abstract}
\citet{Verlinde2016} has recently proposed that spacetime and gravity may emerge from an underlying microscopic theory. In a de Sitter spacetime, such emergent gravity (EG) contains an additional gravitational force due to dark energy, which may explain the mass discrepancies observed in galactic systems without the need of dark matter. For a point mass, EG is equivalent to Modified Newtonian Dynamics (MOND). We show that this equivalence does not hold for finite-size galaxies: there are significant differences between EG and MOND in the inner regions of galaxies. We confront theoretical predictions with the empirical Radial Acceleration Relation (RAR). We find that (i) EG is consistent with the observed RAR only if we substantially decrease the fiducial stellar mass-to-light ratios; the resulting values are in tension with other astronomical estimates; (ii) EG predicts that the residuals around the RAR should correlate with radius; such residual correlation is not observed.
\end{abstract}




\section{Introduction}

Theoretical studies suggest a close relation between black holes and thermodynamics \citep[e.g.,][]{Bardeen1973, Bekenstein1973, Hawking1975}. \citet{Verlinde2011, Verlinde2016} takes the analogy further to suggest that gravity is not a fundamental force of Nature, but emerges from an underlying microscopic theory. This proposal has been named ``emergent'' or ``entropic'' gravity (EG). For a de Sitter spacetime with positive cosmological constant ($\Lambda$), \citet{Verlinde2016} finds that EG contains an additional gravitational force due to $\Lambda$: this ``dark force'' could explain the mass discrepancies observed in galactic systems without the need of particle dark matter. Recent studies investigated EG using weak gravitational lensing \citep{Brouwer2017}, X-ray galaxy clusters \citep{Ettori2016}, dwarf spheroidal galaxies \citep{DiezTejedor2016}, and Solar System tests \citep{Iorio2016}. Here we consider the dynamics of disc galaxies (spirals and irregulars).

In EG the additional force appears below a characteristic acceleration scale $\azero = c H_{0}/6$, where $H_0$ is the Hubble constant and $c$ is the speed of light. This is analogous to Modified Newtonian Dynamics \citep[MOND,][]{Milgrom1983}, but the theoretical basis of the two theories are markedly different \citep[see][for a comment]{Milgrom2016}. Note that \citet{Verlinde2016} uses a slightly different definition of $a_{0}$; here we follow the MOND convention with $\azero=1.2 \times 10^{-10}$ m s$^{-2}$.

MOND dictates that the equations of motion become scale-invariant at accelerations smaller than $a_0$ \citep{Milgrom2009}. This can be achieved through modified laws of gravity \citep{Bekenstein1984, Milgrom2010} or inertia \citep{Milgrom1994}. In modified-inertia theories, test particles on circular orbits obey the following equation:
\begin{equation}\label{eq:MOND}
\gtot = \nu(\gbar/\azero) \cdot \gbar,
\end{equation}
where \gtot\, is the observed ``total'' acceleration, \gbar\, is the Newtonian acceleration from baryonic matter, and $\nu$ is a free function of the theory. Eq.\,\ref{eq:MOND} is also valid in modified-gravity formulations of MOND, but only for systems with 1D symmetry like a spherical galaxy \citep{Bekenstein1984}. $\azero$ and $\nu$ are supposed to be universal in disc galaxies and have been determined empirically by fitting rotation curves \citep{Begeman1991, Sanders2002}. 

\begin{figure*}
\includegraphics[width=0.99\textwidth]{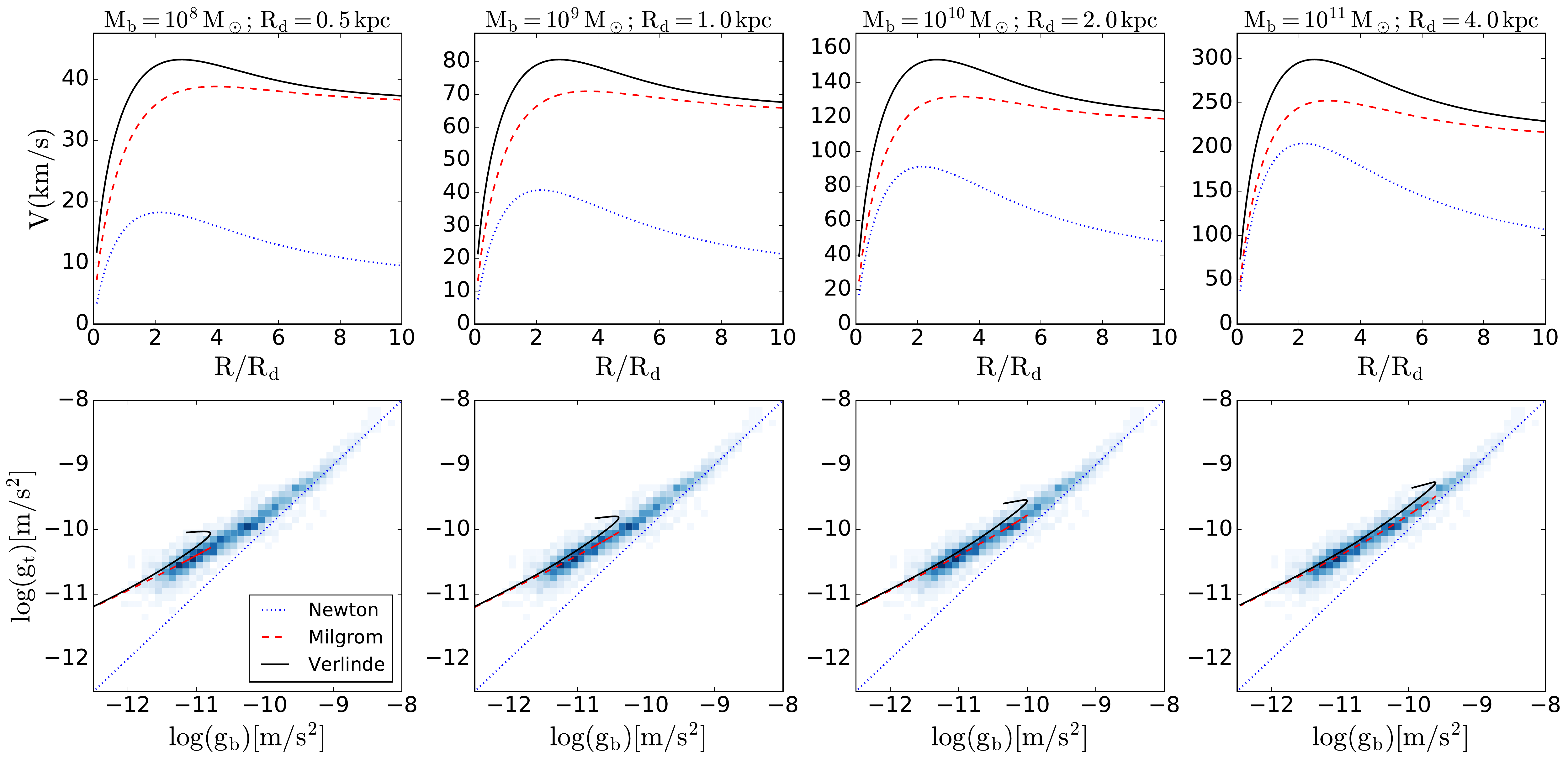}
\caption{Top panels: rotation curves for exponential discs using Newtonian dynamics (dotted line), MOND (dashed line), and EG (solid line). The mass and scale length of each disc model are indicated at the top. Bottom panels: the location of the model galaxies on the radial acceleration relation. The blue scale represents $\sim$2700 datapoints from 153 disc galaxies \citep{McGaugh2016, Lelli2016c}. The lines have the same meaning as in the top panels.}
\label{fig:EGvsMOND}
\end{figure*}
From a pure observational perspective, we have shown that \gbar\ and \gtot\ are tightly correlated in galaxies \citep{McGaugh2016, Lelli2016c}. Specifically, we have built the $Spitzer$ Photometry and Accurate Rotation Curves (SPARC) database \citep{Lelli2016b}: a sample of nearby disc galaxies (spirals and irregulars) with precise estimates of \gbar\ from $Spitzer$ images at 3.6 $\mu$m and \gtot\ from H{\small I}/H$\alpha$ observations. The same radial acceleration relation (RAR) is followed by $\sim$2700 independent points from 153 SPARC galaxies, spanning $\sim$5 dex in baryonic mass and $\sim$4 dex in baryonic surface density. Other types of galaxies (ellipticals, lenticulars, and dwarf spheroidals) follow the same relation as SPARC galaxies \citep{Lelli2016c}. Hence, the RAR seems to be a universal law for galactic systems.

Here we show that EG leads to the following equation:
\begin{equation}\label{Eq:RAR}
 \gtot = f(\gbar/\azero; r) \cdot \gbar,
\end{equation}
where $f$ is specified by the theory and contains an explicit dependence on galaxy radius. This implies that (i) mass discrepancies are higher in EG than MOND (Figure\,\ref{fig:EGvsMOND}), (ii) the RAR should have some intrinsic thickness and the residuals should correlate with radius (Figure\,\ref{fig:EGvsRAR}).

\section{Results}

\subsection{General Equations}\label{sec:general}

We start from Eq. (7.40) of \citet{Verlinde2016}, which relates the apparent dark mass \MD\ to the baryonic mass \Mbar\ for spherically symmetric systems:
\begin{equation}\label{Eq:Verlinde}
 \int_{0}^{r} \dfrac{G \MD^2 (\hat{r})}{\hat{r}^{2}} d\hat{r} = \Mbar \azero r.
\end{equation}
If we differentiate Eq.\ref{Eq:Verlinde} with respect to $r$ and multiply both terms by $G/r^2$, we can easily derive the gravitational acceleration from the apparent dark component:
\begin{equation}
 \gD = \dfrac{G\MD}{r^2}= \sqrt{\azero} \sqrt{\gbar + \dfrac{G}{r}\dfrac{\partial \Mbar}{\partial r}}.
\end{equation}
Hence, the total centripetal acceleration is given by
\begin{equation}\label{Eq:Fede}
\begin{split}
\gtot &= \gbar+ \gD = \gbar \bigg( 1+ \sqrt{\dfrac{\azero}{\gbar}}\sqrt{1 + \dfrac{G}{\gbar r}\dfrac{\partial \Mbar}{\partial r}} \bigg).
\end{split}
\end{equation}

\citet{Verlinde2016} considers the case of a point mass distribution where $\partial \Mbar/\partial r = 0$. When $\gbar\ll\azero$, one recovers the well-known equation for the deep MOND regime:
\begin{equation}\label{eq:DeepMOND}
\gtot = \dfrac{V^2}{r} = \sqrt{\azero\gbar }.
\end{equation}
This naturally explains the observed baryonic Tully-Fisher relation: $V^{4} = \azero G \Mbar$ \citep{McGaugh2000, Lelli2016a}. In actual galaxies, the point-mass approximation is sensible only at very large radii. In the following, we consider a more realistic case where the baryonic mass distribution is extended and symmetric around the center. In such cases, \Mbar\, is a monotonically increasing function of $r$, reaching an asymptotic limit in order to give a finite mass. Hence, $\partial\Mbar/\partial r$ is always positive and tends to zero for large $r$. Clearly, Eq.\,\ref{Eq:Fede} can be generalized by Eq.\,\ref{Eq:RAR}.

\begin{figure*}
\includegraphics[width=0.99\textwidth]{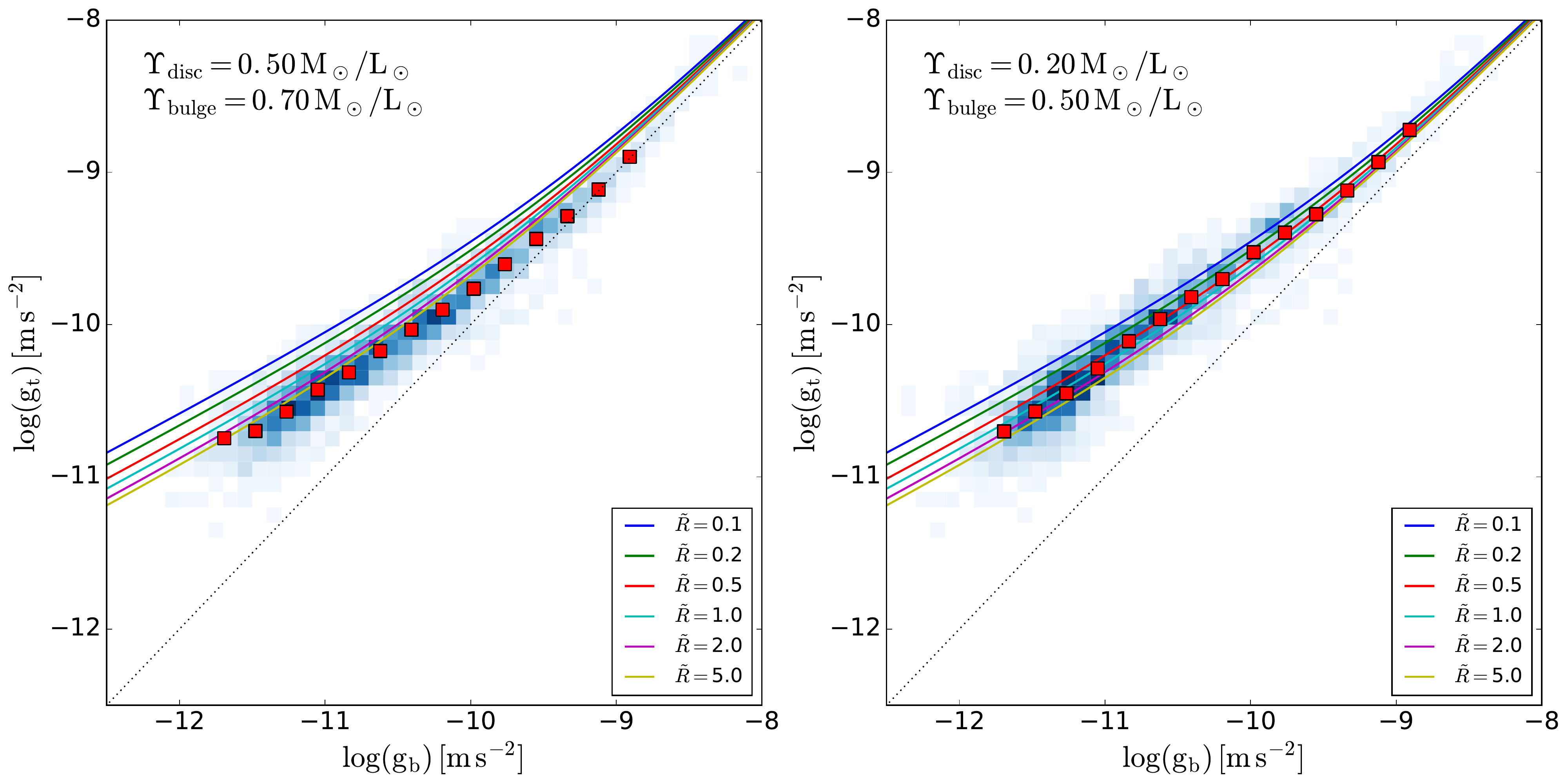}
\caption{Comparison between the observed RAR (blue scale) and the EG predictions for exponential discs (lines). Different lines show Eq.\,\ref{eq:ExpEG} for different values of $\tilde{R} = R/(2R_{\rm d})$. The red squares show the mean of binned data. The left panel shows the RAR for fiducial values of $\Upsilon_{\star}$ at 3.6 $\mu$m from stellar population synthesis models. The right panel shows the RAR adjusting $\Upsilon_{\star}$ to accommodate EG predictions.}
\label{fig:EGvsRAR}
\end{figure*}
\subsection{Disc Models}\label{sec:discs}

Eq.\,\ref{Eq:Fede} could be tested by fitting individual disc galaxies, where $\gtot$, \gbar\, and $\partial\Mbar/\partial r$ are known from observations \citep[e.g.,][]{Lelli2016b}. Here, we limit ourselves to a general analysis using the RAR. A statistical analysis has the advantage that observational uncertainties on single objects (like the assumed distance and inclination) are averaged out. 

We stress that Eq.\,\ref{Eq:Fede} is formally valid for spherically symmetric systems. In Newtonian dynamics \citep{Binney1987} as well as in MOND \citep{Brada1995}, the difference between spherical and disc geometry leads to corrections of the order of $\sim$20$\%$ in the inner regions. While important for detailed fits, it is sensible to use Eq.\ref{Eq:Fede} as a starting point in EG for a statistical study.

The surface density distribution of disc galaxies can be generally described by exponential profiles. Deviations from pure exponentials are common \citep[e.g.,][]{Lelli2016b}, but we will neglect them for the sake of simplicity. The baryonic mass is then given by
\begin{equation}\label{Eq:Exp}
 M_{\rm b} (<R) = 2 \pi \Sigma_{0}\int_{0}^{R} \exp(-\hat{R}/R_{\rm d}) \hat{R} d\hat{R},
\end{equation}
where the central surface density $\Sigma_{0}$ and the characteristic radius $R_{\rm d}$ can be derived from the observations. Note that we are now working in cylindrical coordinates (as appropriate for disc galaxies). The gravitational force of a razor-thin exponential disc is given by:
\begin{equation}\label{Eq:Freeman}
 \gbar(R; R_{\rm d}, \Sigma_{0}) = 2 \pi G \Sigma_0 \tilde{R}[I_0(\tilde{R}) K_0(\tilde{R}) - I_1(\tilde{R}) K_1(\tilde{R})],
\end{equation}
where $\tilde{R} = R/(2R_{\rm d})$ and $I_n$ and $K_n$ are modified Bessel functions of the first and second kind \citep{Freeman1970}. Combining Eq.\,\ref{Eq:Fede}, Eq.\,\ref{Eq:Exp}, and Eq. \ref{Eq:Freeman}, we derive
\begin{equation}\label{eq:ExpEG}
 \gtot = \gbar \bigg\{ 1 + \sqrt{\dfrac{\azero}{\gbar}}\sqrt{1 + \dfrac{\exp(-2\tilde{R}) }{\tilde{R}[I_0(\tilde{R}) K_0(\tilde{R}) - I_1(\tilde{R}) K_1(\tilde{R})]}}\bigg\}
\end{equation}
When $\gbar\gg\azero$, the second term tends to zero and one recovers $\gtot = \gbar$. When $\gbar\ll\azero$, one obtain the deep-MOND limit $\gtot = \sqrt{\gbar \azero}$ only at large radii. This is guaranteed because $\exp(-2\tilde{R})$ decreases faster than the denominator which goes as $\tilde{R}^{-2}$ (the monopole term dominates the gravitational potential). However, it is not guaranteed at small and intermediate radii. Consequently, Verlinde's theory may be observationally distinguishable from MOND.

Figure\,\ref{fig:EGvsMOND} (top panels) shows predicted rotation curves for four model galaxies with typical values of $M_{\rm b}$ and $R_{\rm d}$. EG (solid lines) predicts larger mass discrepancies in the inner regions than MOND (dashed lines) because it converges more gradually to the Newtonian regime (dotted lines). In particular, massive disc galaxies should be sub-maximal in EG. The predicted rotation curves in EG show a prominent decline at intermediate radii and are less flat than in the MOND case. Declining rotation curves are often observed in massive spirals with large bulges (Sa to Sb), but are very rare in low-mass galaxies for which exponential discs are a better description of the stellar mass distribution. For low-mass galaxies, however, one should also account for the gas, which often dominates the mass budget and is more extended than the stellar component \citep[e.g.,][]{Lelli2016b}. This is beyond the scope of our current discussion.

\subsection{Comparison with the observations}\label{sec:RAR}

Figure\,\ref{fig:EGvsMOND} (bottom panels) shows the location of model galaxies on the RAR. In MOND, model galaxies lie exactly on the mean relation as long as Eq.\,\ref{eq:MOND} is assumed and some appropriate $\nu$ is chosen. We adopt the ``simple'' function $\mu(x) = x/(1+x) = \nu^{-1}(y)$, where $x=\gtot/\azero$ and $y=\gbar/\azero$ \citep{Famaey2005}. In EG, model galaxies converge to the mean RAR at large radii, but display a ``hook'' shape above the mean RAR at small radii. This ``hook'' results from the additional radial dependence in Eq.\,\ref{Eq:RAR} and occurs at $R\simeq R_{\rm d}$. The difference for a single galaxy may seem small, but it is systematic.

In Figure\,\ref{fig:EGvsRAR} (left), we confront EG with the data by plotting Eq.\,\ref{eq:ExpEG} for different values of $\tilde{R}$. The predictions from EG lie systematically above our fiducial RAR at either high \gbar\ or small $\tilde{R}$. We recall, however, that the detailed shape of the RAR depends on the assumed stellar mass-to-light ratios ($\Upsilon_{\star}$) for the disc and bulge components, which are used to compute \gbar\ \citep[see Sect.\,4.1 of][]{Lelli2016c}. 

In Figure\,\ref{fig:EGvsRAR} (right), we tune the values of $\Upsilon_{\star}$ to accommodate the predictions of EG. We find that both $\Upsilon_{\rm bulge}$ and $\Upsilon_{\rm disc}$ must be significantly decreased. The resulting values cannot be ruled out, but are in tension with other astronomical estimates (as we discuss in Sect.\,\ref{sec:Disc}). Moreover, we note that this tuning exercise becomes less effective at $\gbar \lesssim 10^{-11}$ m~s$^{-2}$ because the gas contribution starts to dominate and the values of $\gbar$ are largely insensitive to the assumed $\Upsilon_{\star}$.

Independently of the value of $\Upsilon_{\star}$, EG predicts that the RAR should have some intrinsic thickness and the residuals around the mean relation should correlate with radius. The existing data suggests that the RAR is consistent with no intrinsic scatter and there is no significant correlation between residuals and radius (see Figure 4 in \citealt{Lelli2016c}).

\section{Discussion \& Conclusions}\label{sec:Disc}

In this letter, we build simple models for disc galaxies in EG \citep{Verlinde2016} and compare them with the observed RAR \citep{McGaugh2016, Lelli2016c}. We find that EG is consistent with the data only if we change our fiducial conversions from $Spitzer$ [3.6] luminosity to stellar mass. EG implies that $\Upsilon_{\rm disc} \simeq 0.2$ $M_{\odot}/L_{\odot}$ and $\Upsilon_{\rm bulge} \simeq 0.5$ $M_{\odot}/L_{\odot}$ on average, which are significantly smaller than our fiducial values ($\Upsilon_{\rm disc} \simeq 0.5$ $M_{\odot}/L_{\odot}$ and $\Upsilon_{\rm bulge} \simeq 0.7$ $M_{\odot}/L_{\odot}$).

Using stellar population synthesis models, different groups consistently find that the mean $\Upsilon_{\star}$ in the near infrared is somewhat between $\sim$0.4 and $\sim$0.6 $M_{\odot}/L_{\odot}$ for galaxy discs \citep{McGaugh2014, Meidt2014, Schombert2014, Querejeta2015, Herrmann2016, Norris2016}. In principle, values as low as $\sim$0.2 $M_{\odot}/L_{\odot}$ can be derived if one substantially changes the stellar initial mass function and/or the modelling of asymptotic giant branch stars. These possibilities cannot be ruled out, but seem unlikely.

Hydrodynamical models of gas flows in massive spirals suggest that stellar discs are nearly maximal \citep{Kranz2001, Kranz2003, Weiner2001, Weiner2004, Perez2004, Zanmar2008, Fragkoudi2017}, corresponding to $\Upsilon_{\star} \simeq 0.5-0.6$ $M_{\odot}/L_{\odot}$ in the near infrared. Moreover, these values of $\Upsilon_{\star}$ provide sensible trends between the gas fraction and the Hubble type, in line with density wave theory \citep{Lelli2016b}.

The baryonic Tully-Fisher relation (BTFR) also provides important constrains on $\Upsilon_{\star}$ \citep{McGaugh2015, Lelli2016a}. Mean values of $\Upsilon_{\star}\lesssim 0.4$ $M_{\odot}/L_{\odot}$ would significantly increase the BTFR scatter and imply BTFR slopes shallower than 3.5 \citep[see Fig.1 in][]{Lelli2016a}. Hence, EG would not be self-consistent with all the observational constrains, since it predicts a BTFR with nearly zero intrinsic scatter and slope of 4 (like MOND).

Intriguingly, the DiskMass survey measures the vertical velocity dispersion of disc stars in face-on galaxies and finds sub-maximal discs \citep{Bershady2011, Martinsson2013}, corresponding to $\Upsilon_{\rm disc}\simeq0.2$ $M_{\odot}/L_{\odot}$ at [3.6]. This method requires strong assumptions on the disc vertical scale height: different assumptions would lead to nearly maximal discs \citep[see][]{Aniyan2016}.

Finally, in the EG framework, the ``effective'' interpolation function $f$ has an explicit dependence on $R$ (see Eq.\,\ref{Eq:RAR}). This implies that the RAR should display some intrinsic scatter and the residuals should correlate with radius. The current data, instead, are consistent with zero intrinsic scatter and no residual dependences \citep{Lelli2016c}. Larger galaxy samples and more precise observations would help to better understand the intrinsic scatter around the RAR.

We conclude that the observed dynamics of disc galaxies pose a significant challenge to EG. While the asymptotic behaviour is correct in the point mass limit, problems appear for galaxies of finite extent.

\section*{Acknowledgements}

We thank Erik Verlinde for clarifications about his theory proposal. This publication was made possible through the support of the John Templeton Foundation. The opinions expressed here are those of the authors and do not necessary reflect the views of the John Templeton Foundation.




\bibliographystyle{mnras}
\bibliography{EGvsRAR}

\bsp	
\label{lastpage}
\end{document}